\documentclass{goeproc}
\makeindex
\bibpunct{[}{]}{,}{n}{}{,}
 \index{solid state physics, organic solids, organic conductor, low-dimensional metals, charge order, collective excitations, high-frequency measurements, microwave experiments, optical properties}
\begin{document}

\title{Charge-Ordering Phenomena\newline
 in One-Dimensional Solids}

\author{Martin Dressel}

\affil{1. Physikalisches Institut, Universit\"at Stuttgart, \newline Pfaffenwaldring 57, 70550 Stuttgart, Germany}
\affil{\textit{Email:dressel@pi1.physik.uni-stuttgart.de} }

\runningtitle{Charge Order in One-Dimensional Solids}
\runningauthor{Martin Dressel}

\firstpage{1}

\maketitle

\begin{abstract}
As the dimensionality is reduced, the world becomes more and more
interesting; novel and fascinating phenomena show up which call for understanding. Physics in one dimension
is a fascinating topic for theory and experiment: for the former often a simplification, for the latter always a challenge. Various ways
will be demonstrated how one-dimensional structures can be achieved in reality. In
particular organic conductors could establish themselves as model
systems for the investigation of the physics in reduced
dimensions; they also have been subject of intensive research at
the Dritte Physikalische Institut of G\"ottingen University over
several decades.

In the metallic state of a one-dimensional solid, Fermi-liquid
theory breaks down and spin and charge degrees of freedom become
separated. But the metallic phase is not stable in one dimension:
as the temperature is reduced, the electronic charge and spin tend
to arrange themselves in an ordered fashion due to strong
correlations. The competition of the different interactions is
responsible for which broken-symmetry ground state is eventually
realized in a specific compound and which drives the system
towards an insulating state.

Here we review the various ordering phenomena and how they can be identified by dielectric and optic measurements. While the final results
might look very similar in the case of a charge density wave and a charge-ordered metal, for instance, the physical cause is completely different. When density waves form, a gap opens in the electronic density-of-states at the Fermi energy due to nesting of the one-dimension Fermi-surface sheets.
When a one-dimensional metal becomes a charge-ordered Mott insulator, on the other hand, the
short-range Coulomb repulsion localizes the charge on the lattice sites and even causes certain charge patterns.
\end{abstract}


\section{Introduction}
Although the world is three-dimensional in space, physics in one dimension has always attracted a lot of attention. One-dimensional models are simpler compared to three-dimensional ones, and in many cases can be solved analytically only then \cite{LiebMattis66}. Often the reduction of dimension does not really matter because the essential physics remains unaffected. But there are also numerous phenomena in condensed matter which only or mainly occur in one dimension.
In general, the dominance of the lattice is reduced and electronic interactions become superior. Quantum mechanical effects are essential as soon as the confinement approaches the electronic wavelength. Fundamental concepts of physics, like the Fermi liquid theory of interacting particles break down in one dimension and have to be replaced by alternative concepts based on collective excitations \cite{Giamarchi04}.
The competition of different interactions concerning the charge, spin, orbital and lattice degrees of freedom can cause ordering phenomena, i.e.\ phase transitions to a lower-symmetry state as a function of temperature or some order parameter. In one dimension, fluctuations strongly influence the physical properties and smear out phase transitions.
An interesting task now is to approximate one-dimensional systems in reality and check the theoretical predictions. Besides pure scientific interest, the crucial importance of these phenomena in nanotechnology might not lie too far ahead.

\section{Realization of One-Dimensional Systems}
\subsection{Artificial Structures}
The ideal one-dimensional system would be an infinite chain of
atoms in vacuum; close enough to interact with their neighbors,
but completely isolated from the environment. Over the past years,
significant progress has been made towards the realization of
one-dimensional atomic gases, based on Bose-Einstein condensates
of alkalides trapped in two-dimensional optical lattices
\cite{Moritz03}; however, besides other severe drawbacks, only a
limited number of investigations can be performed on quantum gases
in order to elucidate their properties.

\begin{figure}[bp]
\centering\includegraphics[width=10cm]{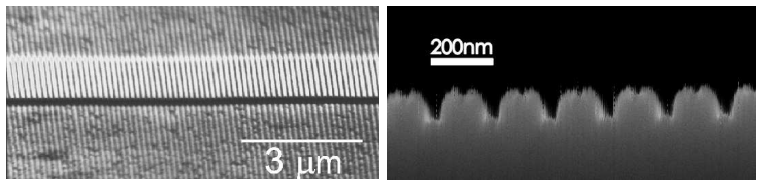}
\caption{\label{fig:quantumwires} One-dimensional semiconductor
quantum wells for GaN lasers (electron micrographs provided by H.
Schweizer, Stuttgart). (a) The ridge waveguide covers an area of
$1000\times 6~\mu$m; (b) the second order grating has a period of
190~nm. }
\end{figure}
In solids one-dimensional physics can be achieved in various ways.
The most obvious approach would be to utilize semiconductor
technology. There layers can be prepared by atomic precision,
using molecular beam epitaxy that leads to a two-dimensional
electron gas at interfaces and quantum wells \cite{Davies98}.
Employing electron-beam lithography and advanced etching
technology, one-dimensional quantum wires are fabricated with an
effective width comparable to the wavelength of the electrons
(Fig.~\ref{fig:quantumwires}). Besides the enormous technological
effort, the disadvantage of this approach is that these structures
are embedded in bulk materials and not easily accessible to
further experiments.

\begin{figure}
\centering\includegraphics[width=10cm]{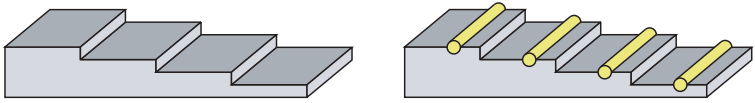}
\caption{\label{fig:goldwires} Realization of metallic nanowires:
The silicon surface is cut in a certain angle leading to single
atomic steps; the width of the terrace depends on the angle.
Evaporated gold assemble itself in such a way that atomic wires
are formed along the steps.}
\end{figure}
If the surface of a single crystal, like silicon, is cut in a
small angle with respect to a crystallographic direction, terraces
are produced on the surface with mono-atomic steps separating them. The surface
reconstruction may lead to an anisotropic arrangement with the
possibility of one-dimensional structures. When a metal, like gold,
is evaporated on top of it, the atoms will organize themselves in
rows along these steps as visualized in Fig.~\ref{fig:goldwires}.
Taking care of the surface reconstruction and of the right density
of gold eventually leads to chains of gold atoms separated by the
terrace width \cite{Himpsel01}. This is a good model of a
one-dimensional metal which can be produced in large quantities,
easily covering an area of $1\times 1$~cm$^2$. As common in
surface technology, ultra-high vacuum is required, and only in
situ experiments -- like electron diffraction, tunnelling or
photoemission spectroscopy--  have been performed by now.

\begin{figure}[h]
\centering\includegraphics[width=6cm]{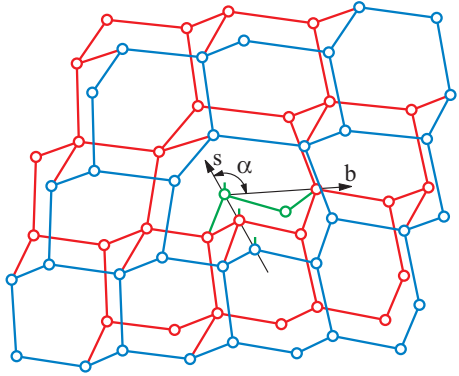}
\caption{\label{fig:dislocations}
 $60^{\circ}$ dislocation in a (111) plane of a diamond lattice, the Burgers vector points in the direction of {\bf b}. At the edge of the additional plane (indicated by {\bf s}) the covalent bonds have no partner.}
\end{figure}

One-dimensional topological defects in single crystals, known as
dislocations, are an intriguing possibility to achieve a
one-dimensional metal, which was utilized by H.-W. Helberg and his
group \cite{Dressel86} in the frame of the G\"ottinger
Sonderforschungsbereich 126. Dislocations in silicon consist of
chains of Si atoms, each having a dangling bond as depicted in
Fig.~\ref{fig:dislocations}, i.e.\ a non-saturated half-filled
orbital \cite{Alexander00}. Along these rows, metallic conduction
is possible while in the perpendicular direction they are
isolated. Since dc measurements with microcontacts on both ends of
a single dislocation are challenging, contactless microwave
experiments were developed as the prime tool to investigate the
electronic transport along dislocations in silicon and germanium \cite{Dressel86}.

It is possible to grow bulk materials as extremely thin and long
hair-like wires when stress is applied; they are known as whiskers
of gold, silver, zinc, tin, etc. Metallic whiskers often lead to
circuit shortages and failures, and are sought to be avoided. An
enormous potential of applications is seen in another sort of
filaments solely consisting of carbon atoms: carbon nanotubes.
They can be considered as rolled-up sheets of graphite, with
electrical properties very much depending on the winding ratio.
Single-wall carbon nanotubes with a small diameter and the right
winding ratio are excellent realizations of one-dimensional
conductors \cite{Connell06}.

\subsection{Anisotropic Crystals}
\begin{figure}[h]
\centering\includegraphics[width=9.5cm]{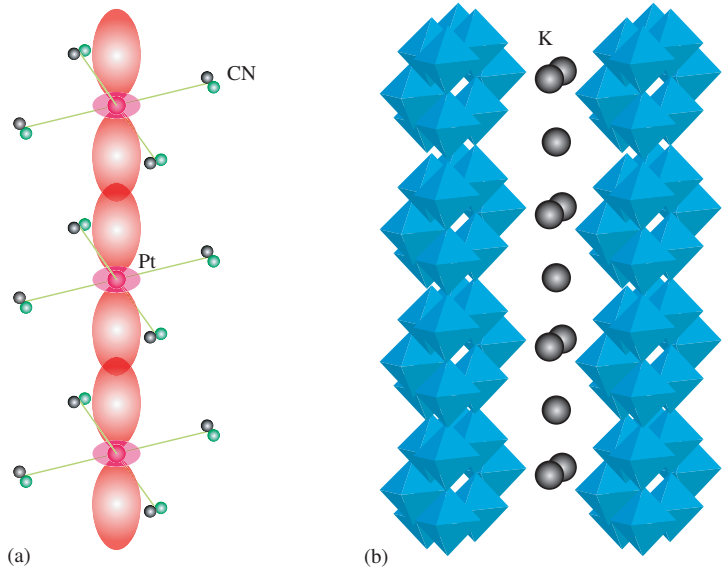}
\caption{\label{fig:KCP1} (a) In
K$_2$Pt(CN)$_4$Br$_{0.3}\cdot$H$_2$O (KCP) the platinum ions form
chains of overlapping orbitals, leading to a metallic
conductivity. (b) Sharing edges and corners, the molybdenum oxide
octahedra in K$_{0.3}$MoO$_3$ (blue bronze) form chains along the
$b$ direction. Alkali-ions like K or Rb provide the charge.}
\end{figure}

By far the most successful approach to one-dimensional physics are
highly anisotropic crystals. Here
K$_2$Pt(CN)$_4$Br$_{0.3}\cdot$H$_2$O, known as KCP, represents the
most intuitive example which consists of a chain of platinum ions
with overlapping $d$ orbitals, as depicted in
Fig.~\ref{fig:KCP1}a. The Pt separation is only 2.894~\AA\ along
the chain direction while the distance between the chains is
9.89~\AA. The Br counterions remove electrons from the planar
Pt(CN)$_4$ units and the resulting fractional charge
Pt$^{1.7}$(CN)$_4$ leads to a partially filled electron band, the
prerequisite for metallic behavior. The room temperature
conductivity along the chain direction is very high
$\sigma_{\parallel} = 10^2~(\Omega{\rm cm})^{-1}$. The anisotropy
ratio is $\sigma_{\parallel}/\sigma_{\perp} = 10^5$
\cite{Bruesch75}.

Transition metal oxides are known for decades to form
low-dimensional crystal structures \cite{Monceau85}. Varying the
composition and structural arrangement provides the possibility to
obtain one- and two-dimensional conductors or superconductors, but
also spin chains and ladders. The interplay of the different
degrees of freedom together with the importance of electronic
correlations makes these systems an almost unlimited source for
novel and exciting phenomena and a challenge for their theoretical
understanding \cite{Cox92}. The blue bronze K$_{0.3}$MoO$_3$ and
related compounds established themselves quickly as model systems
to study electronic properties of quasi-one-dimensional metals
above and below the Peierls transition at $T_{\rm CDW}=185$~K
(Fig.~\ref{fig:KCP1}b).

\begin{figure}
\centering\includegraphics[width=12cm]{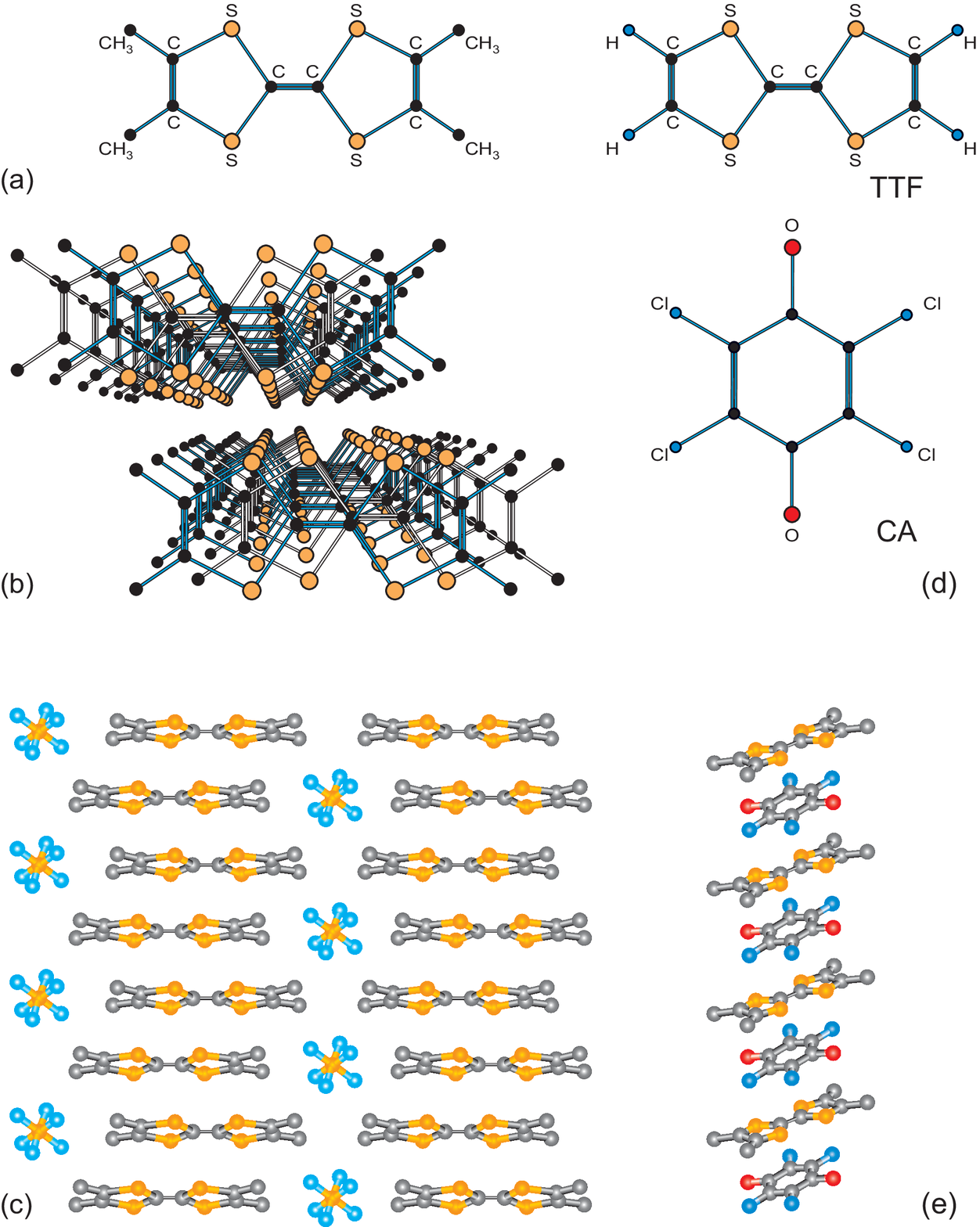}
\caption{\label{fig:structure}
 (a) Planar TMTTF molecule
(b) View along the stacks of TMTTF ($a$-direction) and (c)
perpendicular to them ($b$-direction). Along the $c$-direction the
stacks of the organic molecules are separated by monovalent
anions, like PF$_6^-$ or AsF$_6^-$.\newline (d) TTF molecule and chloranil QCl$_4$ (e) in the
mixed-stack compound TTF-CA, the planar TTF and CA molecules
alternate.}
\end{figure}
While in KCP the metallic properties are due to the platinum ions,
organic conductors form a class of solids with no metal atoms
present (or relevant); instead the $\pi$ electrons distributed
over of the entire organic molecule form the orbitals which might
overlap and lead to band-like conductivity. The additional degree
of freedom, tailoring these molecules, supplements the structural
arrangement in the crystal and makes it possible to fine-tune
competing contributions for the desired properties. This makes organic materials
superior for studying low-dimensional physics and ordering
phenomena in solids. Low-dimensional organic crystals were
explored at the Drittes Physikalisches Institut of G\"ottingen
University already in the 1970s and 1980s; thus in the following
we will constrain ourselves mainly to these examples.

In the course of the last two decades, in particular the Bechgaard
salts
\underline{t}etra\-\underline{m}ethyl-\underline{t}etra\-\underline{s}elena\-\underline{f}ulvalene
(TMTSF), and its variant TMTTF where selenium is replaced by
sulfur, turns out to be an excellent model for
quasi-one-dimensional metals, superconductors, charge order,
spin-density-wave systems, spin chains, spin-Peierls systems,
etc.\ depending on the degree of coupling along and perpendicular
to the chains \cite{Jerome82}. The  planar organic molecules stack
along the $a$-direction with a distance of approximately 3.6~\AA.
In the $b$-direction the coupling between the chains is small, an
in the third direction the stacks are even separated by the
inorganic anion, like PF$_6^-$, SbF$_6^-$, ClO$_4^-$, Br$^-$, etc.
as depicted  in Fig.~\ref{fig:structure}. Each organic molecule
transfers half an electron to the counterions. In general, a small
dimerization leads to pairs of organic molecules. In addition,
spontaneous charge disproportionation, called charge ordering
(CO), may divide the molecules into two non-equivalent species
(cf.\ Fig.~\ref{fig:COpattern}) commonly observed in TMTTF salts.
Due to the instability of the quasi one-dimensional Fermi surface,
at ambient pressure (TMTSF)$_2$PF$_6$ undergoes a transition to a
spin-density-wave (SDW) ground state at $T_{\rm SDW} = 12$~K.
Applying pressure or replacing the PF$_6^-$ anions by ClO$_4^-$
leads to a stronger coupling in the second direction: the material
becomes more two-dimensional. This seems to be a requirement for superconductivity as first discovered in 1979 by J\'erome and coworkers  \cite{Jerome82,Jerome80}.
\begin{figure}
\centering\includegraphics[width=8cm]{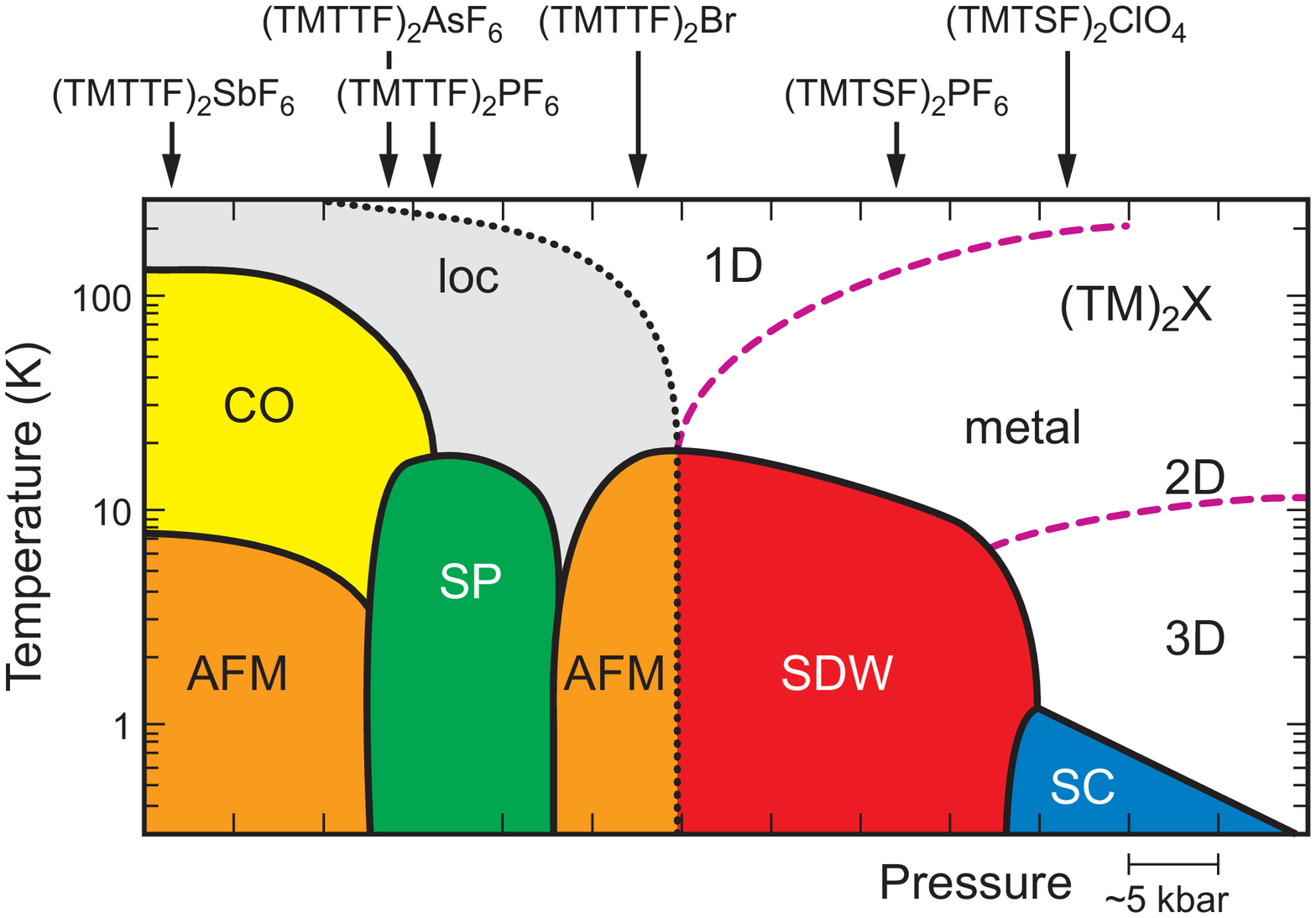}
\caption{\label{fig:phasediagram}
 The phase diagram
of the quasi one-dimensional TMTTF and TMTSF salts. For the different compounds the ambient-pressure position in the phase
diagram is indicated. Going from the left to the right by physical or chemical pressure, the materials get less one-dimensional due to the
increasing interaction in the second and third direction.
 Here \textsf{loc} stands for charge localization,
\textsf{CO} for charge ordering, \textsf{SP} for
spin-Peierls, \textsf{AFM} for antiferromagnet, \textsf{SDW} for spin density wave, and \textsf{SC} for superconductor. The
description of the metallic state changes from a one-dimensional Luttinger liquid to a two- and
three-dimensional Fermi liquid. While some of the boundaries are clear phase transitions,
the ones indicated by dashed lines are better characterized as a crossover.}
\end{figure}

\section{Ordering Phenomena}
One-dimensional structures are intrinsically instable for thermodynamic reasons. Hence various kinds of ordering phenomena can occur which break the translational symmetry of the lattice, charge or spin degree of freedom.
On the other hand, fluctuations suppress long-range order at any finite temperature in one (and two) dimension. Only the fact that real systems consist of one-dimensional chains, which are coupled to some degree, stabilizes the ordered ground state. The challenge now is to extract the one-dimensional physics from experimental investigations of quasi-one-dimensional systems.

At first glance, there seems to be no good reason that in a chain
of molecules the sites are not equivalent, or that the itinerant
charges of a one-dimensional metal are not homogeneously
distributed. However, the translational symmetry can be broken if
electron-phonon interaction and electron-electron interaction
become strong enough. Energy considerations then cause a
redistribution in one or the other way, leading to charge density
waves or charge order. Indeed, these ordering phenomena affect
most thermodynamic, transport and elastic properties of the
crystal; here we want to focus on the electrodynamic response,
i.e.\ optical properties in a broad sense.

First of all, there will be single-particle electron-hole excitations which require energy of typically an eV. But in addition, collective modes are expected. There is a rather general argument by Goldstone \cite{Goldstone61} that whenever a continuous symmetry is broken, long-wavelength modulations in the symmetry direction should occur at low frequencies.
The fact that the lowest energy state has a broken symmetry means that the system is stiff: modulating the order parameter (in amplitude or phase) will cost energy. In crystals, the broken translational order introduces a rigidity to shear deformations, and low-frequency phonons. These collective excitations are expected well below a meV.

\subsection{Charge Density Wave}
\begin{figure}[h]
\centering\includegraphics[width=11cm]{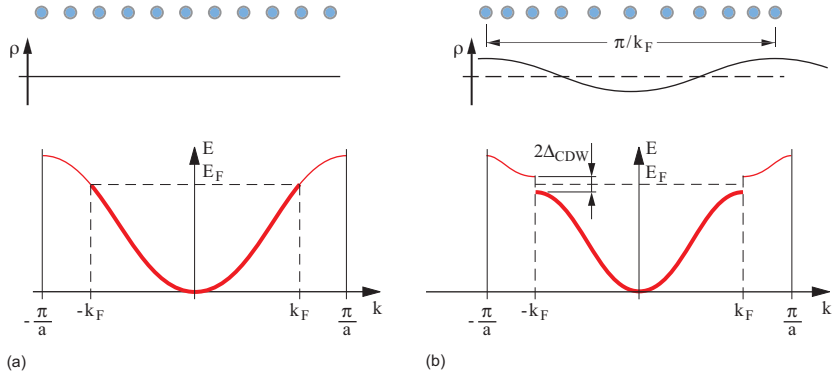}
\caption{\label{fig:cdw1} (a)~In a regular metal, the charge is
homogeneously distributed in space. The conduction band is filled
up to the Fermi energy $E_F$. (b)~A modulation of the charge
density with a wavelength $\lambda=\pi/k_F$ changes the
periodicity; hence in $k$-space the Brillouin zone is reduced
which causes a gap $2\Delta_{\rm CDW}$ at $\pm k_F$. The system
becomes insulating.}
\end{figure}
The energy dispersion forms electronic bands which are filled up
to the Fermi wave-vector $\bf k_F$. In one dimension, the Fermi
surface consists of only two sheets at $\pm k_F$. The crucial
point is that the entire Fermi surface can be mapped onto itself
by a $2k_F$ translation. Since the density of states in one
dimension diverges as $(E-E_0)^{-1/2}$ at the band-edge $E_0$, the
electronic system is very susceptible to $2k_F$ excitations. The
result of the Fermi surface nesting and divergency of the
electronic density of states is a spatial modulation in the charge
density $\rho({\bf r})$ with a period of $\lambda=\pi/k_F$
(Fig.~\ref{fig:cdw1}), which does not have to be commensurate to
the lattice: this is called a charge density wave (CDW).
Long-range charge modulation is crucial because a CDW is a
$k$-space phenomenon. Mediated by electron-phonon coupling, this
causes a displacement of the underlying lattice (Peierls
instability). The gain in electronic energy due to the lowering of
the occupied states has to over-compensate the energy required to
modulate the lattice \cite{Monceau85,Gruner94}.

\begin{figure}[h]
\centering\includegraphics[width=8cm]{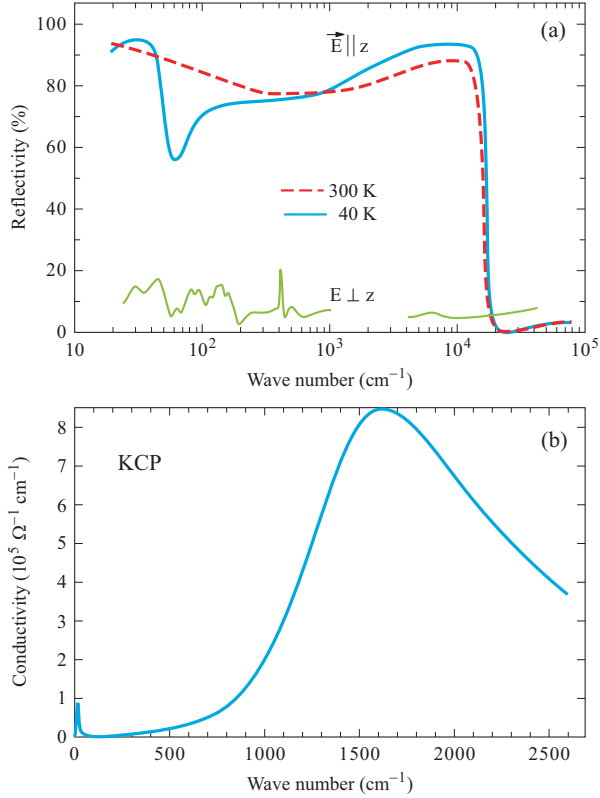}
\caption{\label{fig:cdw2}
 (a) Reflectivity of K$_2$Pt(CN)$_4$Br$_{0.3}\cdot$H$_2$O (abbreviated KCP) measured parallel and perpendicular
to the chains at different temperatures as indicated.
(b) Optical conductivity of KCP for $\bf E\parallel$~stacks at $T=40$~K (after \protect\cite{Bruesch75}). The excitations across the single-particle Peierls gap lead to a broad band in the mid-infrared while the small and sharp peak centered around 15~cm$^{-1}$ is due to the pinned mode.}
\end{figure}
\begin{figure}
\centering\includegraphics[width=8cm]{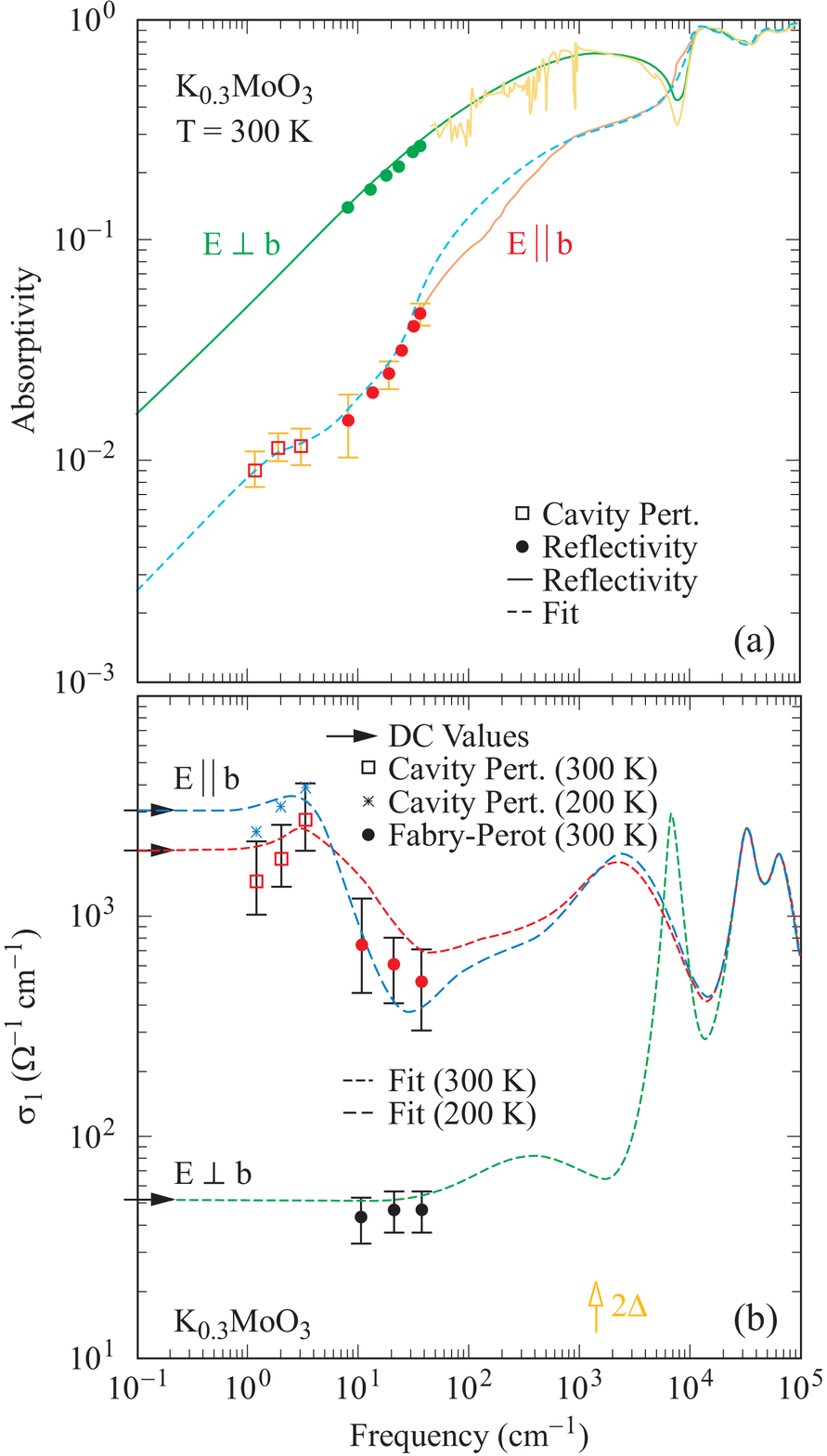}
\caption{\label{fig:cdw4}  (a) Frequency dependence of the room
temperature absorptivity $A=1-R$ of blue bronze (K$_{0.3}$MoO$_3$)
in both orientations $\bf E\parallel$~stacks and $\bf
E\perp$~stacks. The squares were obtained by measuring the surface
resistance using cavity perturbation method, the circles represent
data of quasioptical reflectivity measurements employing a
Fabry-Perot resonator. The solid lines show the results of the
dispersion analysis of the data. (b) Optical conductivity of
K$_{0.3}$MoO$_3$ measured parallel and perpendicular to the stacks
by standard dc technique (arrows), cavity perturbation (open
squares), coherent-source THz spectroscopy (solid dots) and
infrared reflectivity. The open arrow indicates the
single-particle gap as estimated from dc measurements below
$T_{\rm CDW}$ (after \protect\cite{Gorshunov94}).}
\end{figure}

The consequence of the CDW formation is an energy gap $2\Delta$ in the single-particle excitation spectrum, as observed in the activated behavior of electronic transport or a sharp onset of optical absorption.
Additionally, collective excitations are possible which lead to  translation of the density wave as a whole. Although pinning to lattice imperfections prevents Fr\"ohlich superconductivity, the density-wave ground state exhibits several spectacular features, like a pronounced non-linearity in the charge transport (sliding CDW) and a strong oscillatory mode in the GHz range of frequency (pinned-mode resonance) \cite{Gruner94,DresselGruner02}. In 1974 this behavior was observed for the first time in the optical properties of KCP \cite{Bruesch75}, but later recovered in all CDW systems. In Fig.~\ref{fig:cdw2} the optical reflectivity and conductivity of KCP is displayed for different temperatures and polarizations. Due to the anisotropic nature, the reflectivity $R(\omega)$ shows a plasma edge only for the electric field $\bf E$ along the chains while it remains low and basically frequency independent perpendicular to it, as known from dielectrics. At low temperatures, the single particle gap around 1000~cm$^{-1}$ becomes more pronounced, and an additional structure is observed in the far-infrared conductivity which is assigned to the pinned-mode resonance induced by the CDW (Fig.~\ref{fig:cdw2}b).

\begin{figure}[h]
\centering\includegraphics[width=5cm]{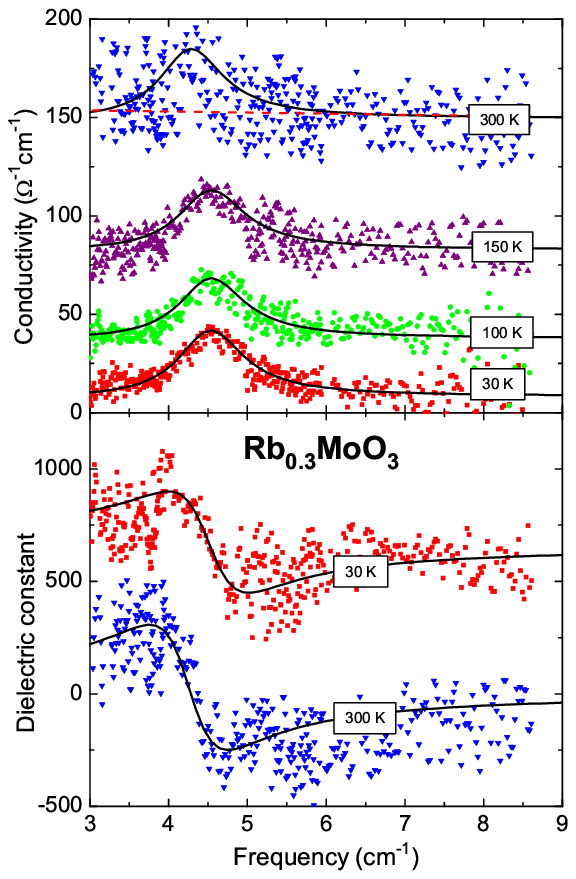}
\caption{\label{fig:cdw3}  Optical conductivity and dielectric
constant of Rb$_{0.3}$Mo$_3$ at various temperatures above and
below $T_{\rm CDW}$ as indicated; note the curves are {\em not}
shifted. The points represent results directly calculated from the
transmission and phase-shift spectra. The solid lines correspond
to fits (after \protect\cite{Pronin98}). Around $\omega_0/2\pi c= 4.5~{\rm cm}^{-1}$ the
pinned-mode resonance is clearly observed which becomes more
pronounced as the temperatures is reduced below $T_{\rm
CDW}\approx 180$~K. The opening of the single particle gap causes
the dielectric constant to increase drastically to approximately
700; the pinned-mode resonance leads to an additional contribution
which is present already at room temperature due to fluctuations.}
\end{figure}

A detailed investigation of the pinned-mode resonance, its center
frequency and lineshape, and furthermore its dependence on temperature
and impurity content turned out to be extremely difficult because it
commonly occurs in the range of 3 to 300 GHz (0.1 to 10~cm$^{-1}$);
i.e.\ it falls right into the gap between high-frequency experiments
using contacts and optical measurements by freely travelling waves
\cite{DresselGruner02}. Microwave technique based on resonant cavities
and quasioptical THz spectroscopy was advanced over the years in order
to bridge this so-called THz gap \cite{Klein93}. Enclosed resonators
have been utilized for decades at the Drittes Physikalisches Institut
\cite{Helberg66} and were readily available when in 1971 I. Shchegolev
suggested them as a tool for investigating small and fragile
low-dimensional organic crystals like TTF-TCNQ \cite{Buranov71}.

The strong influence of fluctuations in one dimension shifts the actual
transition $T_{\rm CDW}$ well below the mean-field value $T_{\rm
MF}=\Delta/1.76 k_B$. This intermediate temperature range $T_{\rm
CDW}<T<T_{\rm MF}$ is characterized by the opening of a pseudogap in
the density of states, i.e.\ a reduced intensity close to the Fermi
energy which is observed in the magnetic susceptibility but not in dc
transport. Optical experiments also see the development of the
pseudogap and indications of the collective mode all the way up to
$T_{\rm MF}$. Utilizing a combination of different methods, the optical
response of K$_{0.3}$MoO$_3$ was measured parallel and perpendicular to
the highly conducting axis; the results for $T=300$~K and 200~K are
displayed in Fig.~\ref{fig:cdw4}. Clearly pronounced excitations are
discovered in the spectra below 50~cm$^{-1}$ for the electric field
$\bf E$ parallel to the  chains, the direction along which the
charge-density wave develops below the Peierls transition temperature
$T_{\rm CDW}$. These excitations are associated with
charge-density-wave fluctuations that exist even at room temperature
and result in a collective contribution to the conductivity. A single
optical experiment finally brought a confirmation of this view:
Fig.~\ref{fig:cdw3} exhibits results of transmission measurements
through thin films of the blue bronze compound Rb$_{0.3}$MoO$_3$ on an
Al$_2$O$_3$ substrate. The transmission coefficient and phase shift
were recorded simultaneously using a Mach-Zehnder interferometer, which
is driven by backward wave oscillators as powerful and tunable sources
and which operates in the THz range of frequencies (30~GHz to 1500~THz,
1 - 50~cm$^{-1}$) \cite{Volkov85}.

\subsection{Charge Order}
\begin{figure}[h]
\centering\includegraphics[width=7cm]{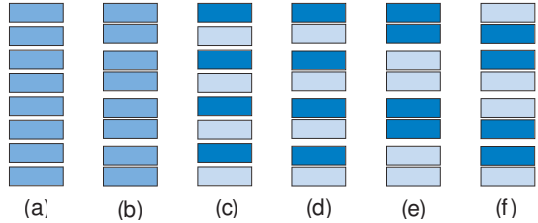}
\caption{\label{fig:COpattern}
 Possible arrangement of the molecules along the stacks. The disproportionation of charge
is depicted by the different gray shade. The molecules can be
dimerized (b), too, which may or may not be accompanied by charge
order (c,d). The periodicity doubles again (teramerization) if
neighboring dimers carry different charge (e), but also if
charge-rich molecules in adjacent dimers form pairs (f).}
\end{figure}

\begin{figure}
\centering\includegraphics[width=6cm]{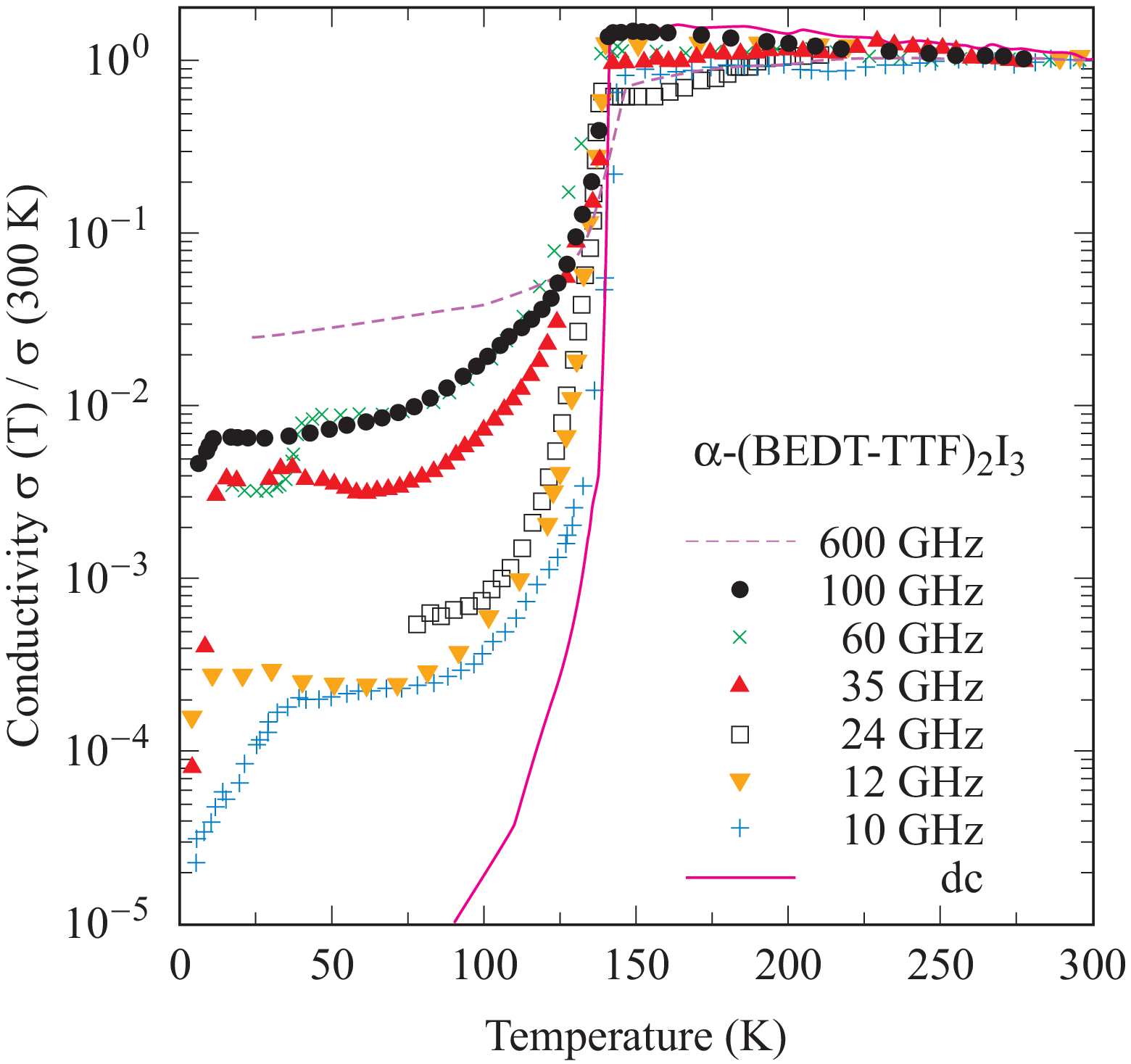}
\caption{\label{fig:aeti}
 Temperature dependent conductivity of $\alpha$-(BEDT-TTF)$_2$I$_3$ within the highly conducting plane measured by dc and microwave technique. The charge-order transition at 135~K leads to a rapid drop of the conductivity. The plateau in the conductivity between 40~K and 100~K increases with frequency indicating hopping conduction (after \protect\cite{Bender84}).}
\vspace*{1cm}
\centering\includegraphics[width=6cm]{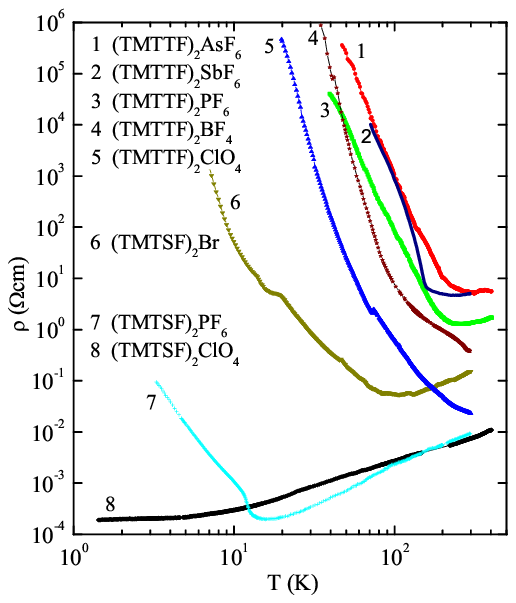}
\caption{\label{fig:dc} Temperature dependence of the dc
resistivity of several Fabre and Bechgaard salts. As the
temperature is reduced, the charges become increasingly localized
in (TMTSF)$_2$AsF$_6$ and (TMTSF)$_2$PF$_6$, before the
charge-ordered state is entered below 100~K. (TMTSF)$_2$SbF$_6$
shows a transition from a metal-like state directly into the
charge-ordered state at $T_{\rm CO}=150$~K. (TMTSF)$_2$PF$_6$
undergoes a SDW transition at $T_{\rm SDW}=12$~K. Only
(TMTSF)$_2$ClO$_4$ remains metallic all the way down to
approximately 1.2~K where it becomes
 superconducting (after \protect\cite{Dressel01}).}
\end{figure}

The crucial point of a CDW is the Fermi surface nesting; the
driving force is the energy reduction of the occupied states right
below the Fermi energy $E_F$ when the superstructure is formed
(cf. Fig.~\ref{fig:cdw1}). Well distinct from a charge density
wave is the occurrence of charge order (CO). The Coulomb repulsion
$V$ between adjacent lattice sites may lead to the preference of
alternatingly more or less charge as depicted in
Fig.~\ref{fig:COpattern}. The extended Hubbard model is a good
description of the relevant energies:
\begin{equation}
{\cal H}= -t\sum_{j=1}\sum_{\sigma=\uparrow\downarrow} \left(c^+_{j,\sigma}c_{j+1,\sigma}+ c^+_{j+1,\sigma}c_{j,\sigma}\right)+
U\sum_{j=1} n_{j\uparrow}n_{j\downarrow} +
V \sum_{j=1}  n_{j} n_{j+ 1} \label{eq:Hubbard} \quad .
\end{equation}
Here $t$ denotes the hopping integral to describe the kinetic
energy, $U$ is the on-site Coulomb repulsion, and $V$ is the
nearest neighbor interaction. The disproportionation of charge on
the molecules represents a short-range order and has to be
commensurate with the lattice.
CO may be accompanied by a slight
lattice distortion (Fig.~\ref{fig:COpattern}d), but this is a
secondary effect. In contrast to a CDW, a metallic state above the
ordering temperature is not required. If it is the case (metallic
state), the gap in the density of states due to the superstructure
also causes a metal-insulator transition. The most intriguing
example of a charge-order driven metal-to-insulator transition was
found in the two-dimensional organic conductor
$\alpha$-(BEDT-TTF)$_2$I$_3$, and this kept the community puzzled
for almost twenty years. Below $T_{\rm CO}=135$~K, the dc and
microwave conductivity (first measured in the
group of H.-W.
Helberg) drops many orders of magnitude (Fig.~\ref{fig:aeti}),
but no modification in the lattice is observed \cite{Bender84}.
Only recently it was understood that electronic correlations are
responsible for this phase transition. Optical experiments (Raman
and infrared) reveal a charge disproportionation from half a hole
per molecule above the phase transition to $0.1e$ and $0.9e$ below
$T_{\rm CO}$; for a review see Dressel and Drichko \cite{Dressel04}.

\begin{figure}
\centering\includegraphics[width=11cm]{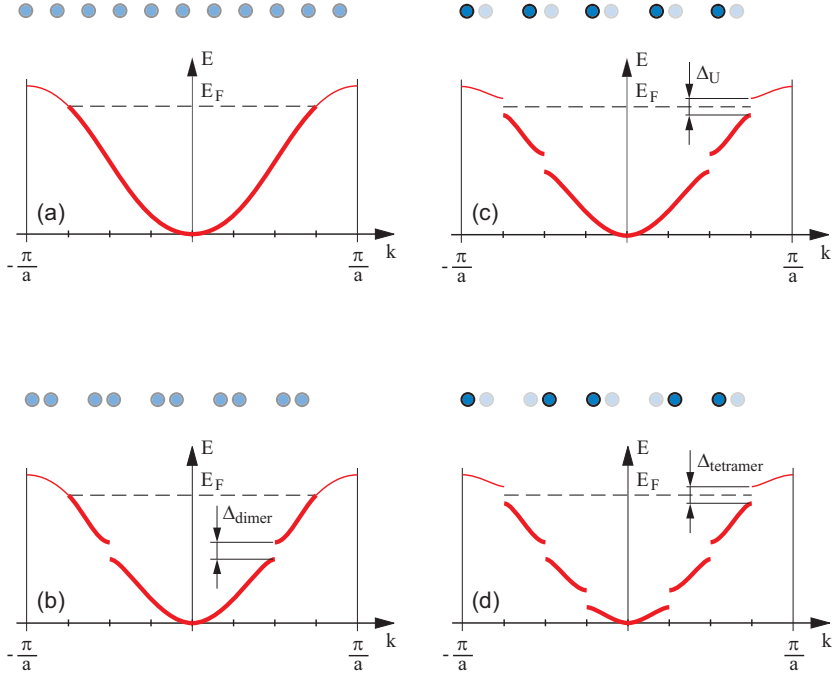}
\caption{\label{fig:co1}(a)~A homogeneous stack of TMT$C$F, for
example, with half an electronic charge $+e$  per molecule results
in a three-quarter-filled band which leads to metallic behavior.
(b)~Dimerization doubles the unit cell and the Brillouin zone is
cut into two equal parts. The upper band is half filled and the
physical properties remain basically unchanged. (c)~Due to on-site
Coulomb respulsion $U$ a gap $\Delta_{\rm U}$ opens at the Fermi
energy $E_F$ that drives a metal-to-insulator transition. (d)~The
tetramerization doubles the unit cell again and also causes a gap $\Delta_{\rm tetramer}$.}
\end{figure}

Similar phenomena can also be observed in the
quasi-one-dimensional (TMTTF)$_2X$ salts which are poor conductors
at ambient temperature and exhibit a rapidly increasing
resistivity as the temperature is lowered (Fig.~\ref{fig:dc}). The
reason is the accumulation of two effects which severely influence
the energy bands as depicted in Fig.~\ref{fig:co1}. The first one
is a structural: due to the interaction with the anions
(Fig.~\ref{fig:structure}c) the molecular stack is dimerized as
visualized in Fig.~\ref{fig:COpattern}b. The conduction band is
split by a dimerization gap $\Delta_{\rm dimer}$ and the material
has a half-filled band. In a second step the Coulomb repulsion $V$
causes charge disproportionation within the dimers
(Fig.~\ref{fig:COpattern}d). On-site Coulomb repulsion $U$ also
drives the one-dimensional half-filled system towards an
insulating state: correlations induce a gap $\Delta_{\rm U}$ at
the Fermi energy $E_F$ as shown in Fig.~\ref{fig:co1}c. The
tetramerization of the CO according to Fig.~\ref{fig:COpattern}e
and f changes this picture conceptually (Fig.~\ref{fig:co1}d): the
soft gap $\Delta_{\rm CO}$ due to short-range nearest-neighbor
interaction $V$ localizes the charge carriers. If not completely
developed it just results in a reduction of the density of state
(pseudogap). The tetramerization gap, on the other hand, is
related to long-range order.

One- and two-dimensional NMR spectroscopy demonstrated the
existence of an intermediate charge-ordered phase in the TMTTF
family. At ambient temperature, the spectra are characteristic of
nuclei in equivalent molecules. Below a continuous charge-ordering
transition temperature $T_{\rm CO}$, there is evidence for two
inequivalent molecules with unequal electron densities. The
absence of an associated magnetic anomaly indicates only the
charge degrees of freedom are involved and the lack of evidence
for a structural anomaly suggests that charge-lattice coupling is
too weak to drive the transition \cite{Chow00}.

\begin{figure}
\centering\includegraphics[width=6cm]{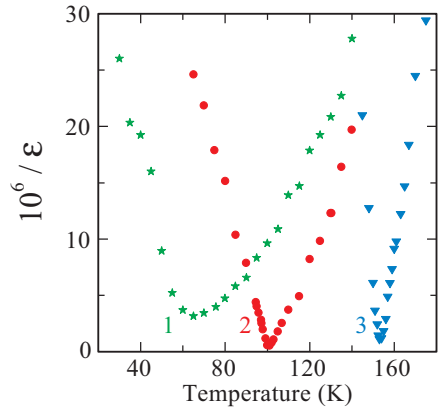}
\caption{\label{fig:TMTTFepsilon} Temperature dependence of the
inverse dielectric constant $1/\epsilon$ of (TMTTF)$_2X$, with
different anions $X$ = PF$_6$ (1), AsF$_6$ (2), and SbF$_6$ (3)
(after \protect\cite{Monceau01}).}
\end{figure}

The
first indications of CO came from dielectric measurements in the
radio-fre\-quen\-cy range \cite{Monceau01}, where a divergency of the
dielectric constant was observed at a certain temperature $T_{\rm
CO}$, as depicted in Fig.~\ref{fig:TMTTFepsilon}. Since this behavior is well known from ferroelectric transitions, the idea is that at elevated temperatures the molecules carry equivalent
charge of $+0.5e$; but upon lowering the temperature, the charge
\begin{figure}[h]
\centering\includegraphics[width=10.5cm]{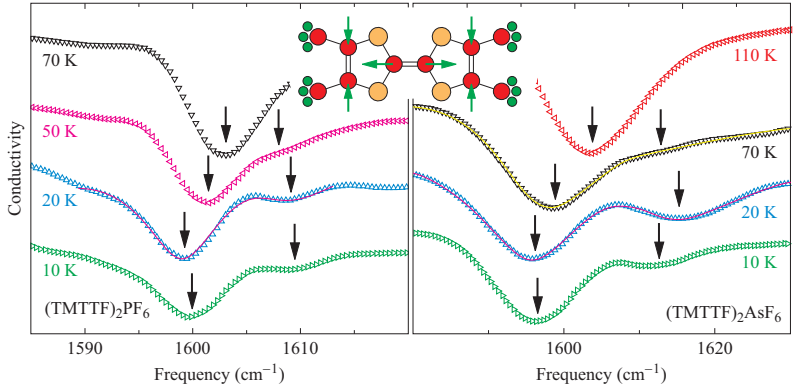}
\caption{\label{fig:CO3} Mid-infrared conductivity of
(TMTTF)$_2$PF$_6$ and (TMTTF)$_2$AsF$_6$ for light polarized
parallel to the molecular stacks. The emv coupled totally
symmetric intramolecular $\nu_3$(A$_g$) mode (which mainly
involves the C=C double bond) splits due to charge order as the
temperature is cooled below $T_{\rm CO}$. The charge
disproportionation ratio amounts to about 2:1 in
(TMTTF)$_2$AsF$_6$ and 5:4 (TMTTF)$_2$PF$_6$ (after \protect\cite{Dumm05}). }
\end{figure}
alternates by $\pm\rho$ causing a permanent dipole moment. Hence,
new intermolecular vibrations at far-infrared frequencies below
100~cm$^{-1}$ get infrared active along all three crystal axes in
the CO state due to the unequal charge distribution on the TMTTF
molecules. Above the CO transition, these modes, which can be
assigned to translational vibrations of the TMTTF molecules, are
infrared silent but Raman active. By now there are no reports on a
collective excitation which should show up as a low-frequency
phonon.

The CO can be locally probed by intramolecular vibrations. Totally
symmetric ${\rm A}_g$ modes are not infrared active; nevertheless,
due to electron-molecular vibrational (emv) coupling (i.e.\ the
charge transfer between two neighboring organic TMTTF molecules
which vibrate out-of phase), these modes can be observed by
infrared spectroscopy for the polarization parallel to the stacks.
As demonstrated in Fig.~\ref{fig:CO3}, the resonance frequency is
a very sensitive measure of the charge per molecule \cite{Dumm05}.
The charge disproportionation increases as the temperature drops below $T_{\rm CO}$ in a mean-field fashion expected from a second-order transition; the ratio amounts to about 2:1 in (TMTTF)$_2$AsF$_6$ and 5:4 (TMTTF)$_2$PF$_6$.
The charge disproportionation is slightly reduced in the AsF$_6$ salt, when it enters the spin-Peierls state, and unchanged in the antiferromagnetic PF$_6$ salt which infers the coexistence of charge order and spin-Peierls order at low temperatures.

\subsection{Neutral-Ionic Transition}
\begin{figure}
\centering\includegraphics[width=6cm]{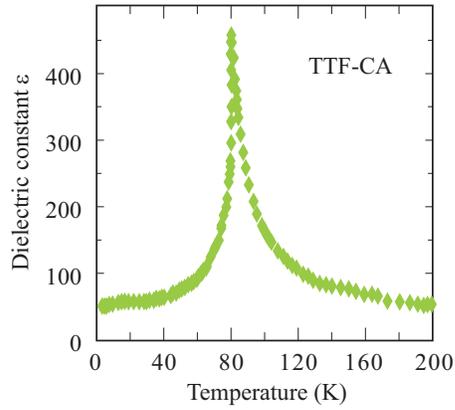}
\caption{\label{fig:TTF-CAepsilon}
 Temperature dependent dielectric constant $\epsilon(T)$ of TTF-CA measured at a frequency of 30 kHz (after \protect\cite{Horiuchi00}). The divergency at $T_{\rm NI}=81$~K clearly evidences the ferroelectric-like neutral-ionic transition.}
\end{figure}

While in the previous example the crystals consist of separate
cation and anion chains between which the electron transfer
occurs, mixed-stack organic charge-transfer compounds have only
one type of chain composed of alternating $\pi$ electron donor and
acceptor molecules (... A$^{-\rho}$D$^{+\rho}$A$^{-\rho}$
D$^{+\rho}$A$^{-\rho}$D$^{+\rho}$ ...) as sketched in
Fig.~\ref{fig:structure}e. These materials are either neutral or
ionic, but under the influence of pressure or temperature certain
neutral compounds become ionic. There is a competition between the
energy required for the formation of a D$^+$A$^-$ pair and the
Madelung energy. Neutral-ionic (NI) phase transitions are
collective, one-dimensional charge-transfer phenomena occurring in
mixed-stack charge-transfer crystals, and they are associated to
many intriguing phenomena, as the dramatic increase in
conductivity and dielectric constant at the transition, such as
plotted in Fig.~\ref{fig:TTF-CAepsilon}
\cite{Torrance81,Horiuchi00}.

In the simplest case the charge per molecule changes from
completely neutral $\rho=0$ to fully ionized $\rho=1$. Ideally
this redistribution of charge is decoupled from the lattice, and
therefore should not change  the inter-molecular spacing. In most
real cases, however, the NI transition is characterized by the
complex interplay between the average ionicity $\rho$ on the
molecular sites and the stack dimerization $\delta$. The ionicity
may act as an order parameter only in the case of discontinuous,
first order phase transitions. While the inter-site Coulomb
\begin{figure}[!ht]
\centering\includegraphics[width=6cm]{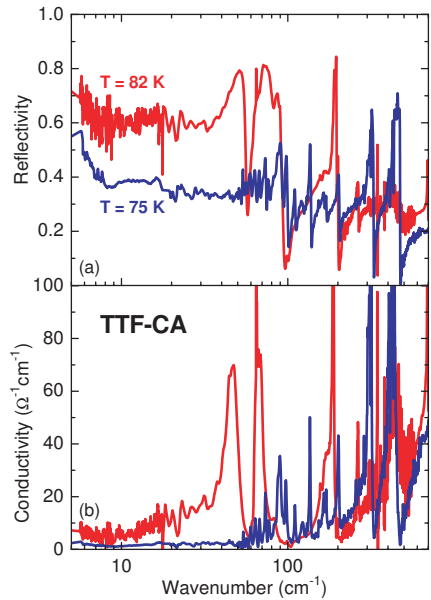}
\caption{\label{fig:TTF-CAref}  (a) Reflectivity and (b)
conductivity spectra of TTF-CA measured along the stacking
direction above (red line) and below (blue line) the neutral-ionic
transition at $T_{\rm NI}=81$~K  (after \protect\cite{Masino06}).}
\end{figure}
interaction $V$ favors a discontinuous jump of ionicity, the
intra-chain charge-transfer integral $t$ mixes the fully neutral
and fully ionic quantum states and favors continuous changes in
$\rho$. The coupling of $t$ to lattice phonons induces the
dimerization of the stack, basically a Peierls-like transition to
a ferroelectric state, which is a second order phase transition.
Intramolecular (Holstein) phonons, on the other hand, modulate the
on-site energy $U$ and favor a discontinuous jump in $\rho$.

In terms of a modified, one-dimensional Hubbard model [similar to
Eq.~(\ref{eq:Hubbard})], the NI transition can be viewed as a
transition from a band insulator to a Mott insulator, due to the
competition between the energy difference between donor and
acceptor sites, and the on-site Coulomb repulsion  $U$. Peierls
and Holstein phonons are both coupled to charge transfer
electrons, albeit before the NI transition the former are only
infrared active, and the latter only Raman active. This makes
polarized Raman and reflection measurements a suitable tool to
explore the NI transition.

The temperature induced NI transition of
tetrathiafulvalene-chloranil (TTF-CA, cf.\ Fig.~\ref{fig:COpattern}d, e) at $T_{\rm NI}=81$~K is the
prime example of a first-order transition with a discontinuous
jump in $\rho$. This can be seen in Fig.~\ref{fig:TTF-CAref}
by a jump in the frequency
of those of the intramolecular vibrations, which are coupled to
the electronic charge because their position depends on the charge on
the molecules \cite{Masino03,Masino06}.

\begin{figure}[h]
\centering\includegraphics[width=6cm]{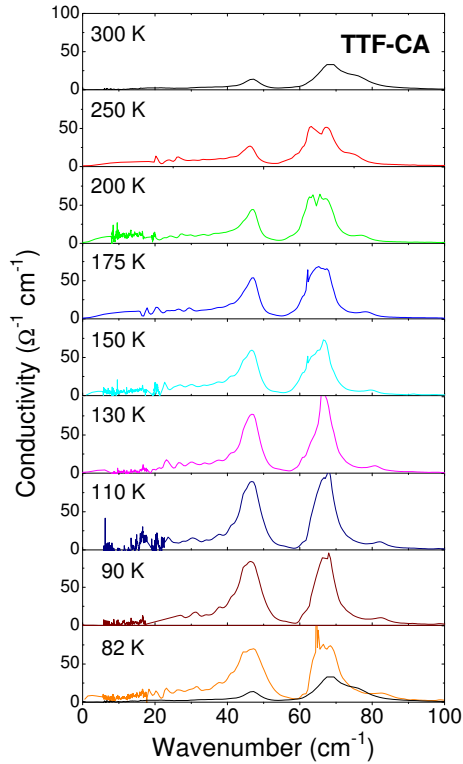}
\caption{\label{fig:TTF-CAcond} Low-frequency conductivity of
TTF-CA for $T>T_{\rm NI}$ for different temperatures as indicated
in the panels. As the NI transition is approached by decreasing
temperature, the modes become stronger and an additional band
appears as low as 20~cm$^{-1}$. To make the comparison easier, the
room temperature spectrum (black line) is replotted in the lowest
frame (after \protect\cite{Masino06}).}
\end{figure}
The vibronic bands present in the infrared spectra for $T>T_{\rm
NI}$ are combination modes involving the lattice mode, which gives rise to the
Peierls distortion at the transition. From calculations we expect
three lattice modes which couple to electrons and become stronger
as the transition is approached. The lattice modes strongly couple
to electrons and behave as soft modes of the ferroelectric
transition at $T_{\rm NI}=81$~K. In Fig.~\ref{fig:TTF-CAcond} the
low-frequency conductivity spectra are plotted for different
temperatures $T>T_{\rm NI}$. The lowest mode softens most and is
seen strongly overdamped around 20~cm$^{-1}$. The temperature
evolution of this Peierls mode, which shows a clear softening
(from 70 to 20~cm$^{-1}$) before the first-order transition to the
ionic ferroelectric state takes place. In the ordered phase, a
clear identification and theoretical modelling of the Goldstone
mode is still an open problem because the system has several
degrees of freedom coupled to each other.

The cooperative charge transfer among the constructive molecules of TTF-CA can also be induced by irradiation of a short laser pulse. A photoinduced local charge-transfer excitation triggers the phase change and cause the transition in both directions \cite{Koshihara99}. When Cl is replaced by Br in the tetrahalo-{\it p}-benzoquinones the lattice is expanded, (like a negative pressure) and the ionic phase vanishes completely. Hydrostatic pressure or Br-Cl substitution is utilized as a control parameter to tune the NI transition more or less continuously at $T\rightarrow 0$ \cite{Horiuchi03}.

\section{Outlook}
No doubt, one-dimensional physics matured from a toy model to an extremely active field of theoretical and experimental research, spanning a broad range from quantum gases to condensed-matter physics and semiconductor technology. Several novel and exciting phenomena can be investigated in these systems. In one-dimensional metals collective modes replace the single-particle excitations common to three-dimensional conductors and  described by Landau's Fermi liquid concept of interaction electrons. Another property typical for low-dimensional solids is their susceptibility to symmetry breaking with respect to the lattice, the charge and the spin degrees of freedom. Broken-symmetry ground states imply that the system becomes stiff, because the modulation of the order parameter costs energy; therefore collective modes appear at low energies. In the case of magnets, the broken rotational symmetry leads to a magnetic stiffness and spin waves. In superconductors the gauge symmetry is broken, but due to the Higgs mechanism the Goldstone mode is absent at low frequencies and shifted well above the plasma frequency. In the examples above, we were dealing with translational symmetry, which is lowered in crystals due to charge ordering phenomena.

Charge density waves drive a metal to an insulator, for the Fermi surface becomes instable; the pinned-mode resonance can nicely be detected in the GHz using a variety of high-frequency and optical techniques. Purely electronic correlations between adjacent sites can cause charge disproportionation. Organic conductors are suitable realizations to investigate the properties at the metal-insulator transitions. The neutral-ionic transition observed in mixed-stack one-dimensional organic charge-transfer salts can be a pure change of ionizity, but commonly goes hand in hand with a Peierls distortion. This can be seen in a softening of the low-frequency phonon modes above the phase transition.

Optical methods and in particular microwave techniques as developed at the Dritte Physikalische Institut in G\"ottingen are powerful tools for investigation of charge-ordering phenomena in solids.

\begin{acknowledgements}
The review is based on many years of collaboration with a large number of people; only some of them can be mentioned here. In particular I would like to thank N. Drichko, M. Dumm, A. Girlando, B. Gorshunov, G. Gr\"uner, and H.-W. Helberg.
\end{acknowledgements}


\begin{thebibliography}{99}
\bibitem{LiebMattis66}{\em Mathematical Physics in One Dimension}, edited by E.H. Lieb and D.C. Mattis (Academic Press, New York,
1966).
\bibitem{Giamarchi04}
T. Giamarchi, {\em Quantum Physics in One Dimension} (Oxford
University Press, Oxford, 2004); M. Dressel, `Spin-charge
separation in quasi one-dimensional organic conductors',
Naturwissenschaften {\bf 90}, 337 (2003).
\bibitem{Moritz03}
H. Moritz, T. St\"oferle, M. K\"ohl, and T. Esslinger, `Exciting
collective oscillations in a trapped 1D gas', Phys. Rev. Lett.
{\bf 91}, 250402 (2003).
\bibitem{Davies98}J.H. Davies, {\em The Physics of Low-Dimensional Semiconductors} (Cambridge University Press, Cambridge, 1998).
\bibitem{Himpsel01}F.J. Himpsel, A. Kirakosian, J.N. Crain, J.-L. Lin, und D.Y. Petrovykh, `Self-assembly of one-dimensional nanostructures at silicon surfaces', Solid State Commun. {\bf 117}, 149 (2001).
\bibitem{Dressel86}M. Dressel and H.-W. Helberg, `AC conductivity of deformed germanium single crystals at $T =4.2$~K',
phys. stat. sol. (a) {\bf 96}, K199 (1986); M. Brohl, M. Dressel,
H.-W. Helberg, and H. Alexander, `Microwave conductivity
investigations in plastically deformed silicon', Phil. Mag. B {\bf
61}, 97 (1990).
\bibitem{Alexander00}
H. Alexander and H. Teichler, {\em Dislocations}, in: {\em Handbook of Semiconductor Technology}, Vol. 1, edited by K.A. Jackson and W. Schr\"oter (Wiley-VCH, New York, 2000), p. 291.

\bibitem{Connell06}M. O'Connell, {\em Carbon Nanotubes} (Taylor \&\ Francis, Boca Raton, 2006); P.J.F. Harris {\em Carbon Nanotubes and Related Structures} (Cambridge University Press, Cambridge, 2004); S. Reich, C. Thomsen, and J. Maultzsch {\em Carbon Nanotubes} (Wiley-VCH, Weinheim, 2004).

\bibitem{Bruesch75}P. Br\"uesch, {\em Optical Properties of the One-Dimensional Pt Complex Compounds}, in: {\em One-Dimensional Conductors}, edited by H.G. Schuster (Springer-Verlag, Berlin, 1975), p.\ 194;
P. Br\"uesch, S. Str\"assler, and H. R. Zeller, `Fluctuations and
order in a one-dimensional system. A spectroscopical study of the
Peierls transition in K$_2$Pt(CN)$_4$Br$_{0.3}\cdot$3(H$_2$O)',
Phys. Rev. B {\bf 12}, 219 (1975).

\bibitem{Monceau85}
{\it Electronic Properties of Inorganic Quasi-One-Dimensional Compounds}, Part I/II, edited by P. Monceau,
(Reidel, Dordrecht, 1985);
{\it Physics and Chemistry of Low Dimensional Inorganic Conductors}, edited by C. Schlenker,
J. Dumas, M. Greenblatt, and S. Van Smalen (Plenum, New York, 1996).
\bibitem{Cox92}P.A. Cox, {\em Transition Metal Oxides} (Clarendon Press, Oxford, 1992);
S. Maekawa, T. Tohyama, S.E. Barnes, S. Ishihara, W. Koshibae, and
G. Khaliullin, {\em The Physics of Transition Metal Oxides}
(Springer, Berlin, 2004).

\bibitem{Jerome82}
D. J{\'e}rome and H.~J. Schulz, `Organic conductors and superconductors',
Adv. Phys. {\bf 31},  299  (1982); D. J\'erome,  in {\em Organic Conductors}, edited by J.-P. Farges (Marcel Dekker, New York, 1994), p.\ 405; M. Dressel, `Spin-charge separation in quasi one-dimensional organic conductors',
Naturwissenschaften {\bf 90}, 337 (2003); M. Dressel, `Ordering phenomena
in  quasi one-dimensional organic conductors', Naturwissenschaften {\bf 94}, DOI 10.1007/s00114-007-0227-1 (2007).
\bibitem{Jerome80}
D. J{\'e}rome, A. Mazaud, M. Ribault, and K.  Bechgaard,
`Superconductivity in a synthetic organic conductor (TMTSF)$_2$PF$_6$,'
J.\ Physique Lett.\  {\bf 41}, L95 (1980).
\bibitem{Goldstone61}J. Goldstone, `Field theories with ``superconductor'' solution', Nuovo cimento {\bf 19}, 154 (1961);
J. Goldstone, A. Salam, and S. Weinberg, `Broken symmetries',
Phys. Rev. {\bf 127}, 965 (1962).


\bibitem{Gruner94}G. Gr\"uner, {\it Density Waves in Solids}, (Addison-Wesley, Reading, MA, 1994).
\bibitem{DresselGruner02}M. Dressel and G. Gr\"uner, {\em Electrodynamics of Solids} (Cambridge University Press, Cambridge, 2002).

\bibitem{Klein93}O. Klein, S. Donovan, M. Dressel, and G. Gr\"{u}ner, `Microwave cavity perturbation technique. Part I:
Principles', Int. J. Infrared and Millimeter Waves {\bf 14}, 2423
(1993); S. Donovan, O. Klein, M. Dressel, K. Holczer, and G.
Gr\"{u}ner, `Microwave cavity perturbation technique. Part II:
Experimental scheme', Int. J. Infrared and Millimeter Waves {\bf
14}, 2459 (1993); M. Dressel, S. Donovan, O. Klein, and G.
Gr\"{u}ner, `Microwave cavity perturbation technique. Part III:
Applications', Int. J. Infrared and Millimeter Waves {\bf 14},
2489 (1993); A. Schwartz, M. Dressel, A. Blank, T. Csiba, G.
Gr\"{u}ner, A.A Volkov, B.P. Gorshunov, and G.V. Kozlov, `Resonant
techniques for studying the complex electrodynamic response of
conducting solids in the millimeter and submillimeter wave
spectral range', Rev. Sci. Instrum. {\bf 66}, 2943 (1995); M.
Dressel, O. Klein, S. Donovan, and G. Gr\"uner, `High frequency
resonant techniques for the study of the complex electrodynamic
response in solids', Ferroelectrics {\bf 176}, 285 (1996).
\bibitem{Helberg66}H.-W. Helberg and B. Wartenberg, `Zur Messung
der Stoffkonstanten $\epsilon$ und $\mu$ im GHz-Bereich mit
Resonatoren', Z. Angew. Phys. {\bf 20}, 505 (1966).
 The tradition goes back to the Institut f\"ur Angewandte Elektrizit\"at (Institute of Applied Electricity) founded in the beginning of the 20th century and headed by Max Reich for a long time. Students like Arthur von Hippel spread this knowledge all around the world and made high-frequency investigations of solids to a powerful tool. The foundation of the {\em Laboratory of Insulation Research} and the {\em Radiation Laboratory} at MIT during World War II certainly had the largest impact.
\bibitem{Buranov71}L.I. Buranov and I.F. Shchegolev, `Method of measuring conductivity of small crystals at a
frequency of 10$^{10}$~HZ', Prib. Tekh. Eksp. (engl.) {\bf 14},
528 (1971); I.F. Shchegolev, `Electric and magnetic properties of
linear conducting chains', phys. stat. sol (a) {\bf 12}, 9 (1972);
H.~W. Helberg and M. Dressel, `Investigations of organic
conductors by the Schegolev method', {J. Phys. I. (France)} {\bf
6}, 1683 (1996).


\bibitem{Gorshunov94}B.P. Gorshunov, A.A Volkov, G.V. Kozlov, L. Degiorgi, A. Blank, T. Csiba, M. Dressel, Y. Kim, A. Schwartz, and G.
Gr\"{u}ner, `Charge density wave paraconductivity in
K$_{0.3}$MoO$_3$', Phys. Rev. Lett. {\bf 73}, 308 (1994); A.
Schwartz, M. Dressel, B. Alavi, A. Blank, S. Dubois, G. Gr\"uner,
B. P. Gorshunov, A. A. Volkov, G. V. Kozlov, S. Thieme, L.
Degiorgi, and F. L\'evy, `Fluctuation effects on the
electrodynamics of quasi one-dimensional conductors above the
charge-density-wave transition', Phys. Rev. B {\bf 52}, 5643
(1995).
\bibitem{Pronin98}A.V. Pronin, M. Dressel, A. Loidl, H.S.J. van der Zant, O.C. Mantel, and C. Dekker,
`Optical investiations of the collective transport in CDW-films',
Physica B {\bf 244}, 103 (1998).
\bibitem{Volkov85}
G. Kozlov and A. Volkov, {\em Coherent Source Submillimeter Wave
Spectroscopy}, in: {\em Millimeter and Submillimeter Wave
Spectroscopy of Solids},  edited by G. Gr\"uner  (Springer,
Berlin, 1998),  p.\ 51; B. Gorshunov, A. Volkov, I. Spektor, A.
Prokhorov, A. Mukhin, M. Dressel, S. Uchida, and A. Loidl,
`Terahertz BWO-spectrosopy', Int.\ J.\ of Infrared and Millimeter
Waves, {\bf 26}, 1217 (2005).

\bibitem{Bender84}K. Bender, K. Dietz, H. Endres, H.-W. Helberg, I. Hennig, H.J. Keller,
 H.W. Sch\"afer, and D. Schweitzer, `(BEDT-TTF)$^{2+}$J$_3^-$ - A two-dimensional organic metal',
 Mol. Cryst. Liq. Cryst. {\bf 107}, 45 (1984);
 M. Dressel, G. Gr\"uner, J.P. Pouget, A. Breining, and D. Schweitzer,
 `Field- and frequency dependent transport in the two-dimensional organic conductor $\alpha$-(BE\-DT\--TTF)$_{2}$\-I$_{3}$'
 J. de Phys. I (France) {\bf 4}, 579 (1994).
\bibitem{Dressel04}M. Dressel and N. Drichko, `Optical properties of two-dimensional organic conductors: signatures
of charge ordering and correlation effects', Chem. Rev. {\bf 104},
5689 (2004);
M. Dressel, `Ordering phenomena
in  quasi one-dimensional organic conductors', Naturwissenschaften {\bf 94}, DOI 10.1007/s00114-007-0227-1 (2007).
\bibitem{Dressel01}
M. Dressel, S. Kirchner, P. Hesse, G. Untereiner, M. Dumm, J. Hemberger,
A. Loidl, and L. Montgomery `Spin and charge dynamics in Bechgaard
salts', Synth. Met. {\bf 120}, 719 (2001).

\bibitem{Chow00}
D.S. Chow,  F. Zamborszky,  B. Alavi, D.J. Tantillo, A. Baur, C.A. Merlic, and S.E. Brown, `Charge ordering in the TMTTF family of molecular conductors',
Phys. Rev. Lett. {\bf 85}, 1698 (2000).

\bibitem{Monceau01}P. Monceau, F. Ya. Nad, and S. Brazovskii,
`Ferroelectric Mott-Hubbard phase of organic (TMTTF)$_2X$
conductors', Phys. Rev. Lett. {\bf 86}, 4080 (2001).

\bibitem{Dumm05}M. Dumm, M. Abaker, and M. Dressel, `Mid-infrared response of charge-ordered quasi-1D organic conductors (TMTTF)$_2X$', J. Phys. IV (France) {\bf 131}, 55
(2005).


\bibitem{Torrance81}J.B. Torrance, J.E. Vazquez, J.J. Mayerle, and V.Y. Lee, `Discovery of a neutral-to-ionic phase transition in organic materials', Phys. Rev. Lett. {\bf 46}, 253 (1981); J.B. Torrance, A. Girlando, J.J. Mayerle, J.I. Crowley, V.Y. Lee, P. Batail,
and S.J. LaPlace, `Anomalous nature of neutral-to-ionic phase
transition in tetrathiafulvalene-chloranil', Phys. Rev. Lett. {\bf
47}, 1747 (1981); T. Mitani, Y. Kaneko, S. Tanuma, Y. Tokura, T.
Koda, and G. Saito, `Electric conductivity and phase diagram of a
mixed-stack charge-transfer crystal:
Tetrathiafulvalene-p-chloranil', Phys. Rev. B {\bf 35}, 427
(1987); S. Horiuchi, Y. Okimoto, R. Kumai, and Y. Tokura, `Anomalous
valence fluctuation near a ferroelectric transition in an organic
charge-transfer complex', J. Phys. Soc. Japan {\bf 69}, 1302
(2000).

\bibitem{Horiuchi00}S. Horiuchi, Y. Okimoto, R. Kumai, and Y. Tokura,
`Anomalous valence fluctuation near a ferroelectric transition in an organic charge-transfer complex', J. Phys. Soc. Jpn. {\bf 69}, 1302 (2000).

\bibitem{Masino03}M. Masino, A. Girlando, and Z.G. Soos, `Evidence for a soft mode in the temperature induced neutral-ionic transition of
TTF-CA', Chem. Phys. Lett. {\bf 369}, 428 (2003).
\bibitem{Masino06}M. Masino, A. Girlando, A. Brillante, R.G. Della Valle, E. Venuti, N. Drichko, and M. Dressel, `Lattice dynamics of TTF-CA across the neutral ionic transition', Chem. Phys. {\bf 325}, 71 (2006);
 N. Drichko {\it et al.}, to be published.
\bibitem{Koshihara99}
S.Y. Koshihara, Y. Takahashi, H. Saki, Y. Tokura, and T. Luty,  `Photoinduced cooperative charge transfer in low-dimensional organic crystals', J. Phys. Chem. B {\bf 103}, 2592 (1999).
\bibitem{Horiuchi03}
S. Horiuchi, Y. Okimoto, R. Kumai, and Y. Tokura,
`Quantum phase transition in organic charge-transfer complexes', Science {\bf 299}, 229 (2003).



\end{thebibliography}
\end{document}